\documentclass[11pt]{aastex}
\usepackage{emulateapj5,apjfonts}
\usepackage{onecolfloat}
\usepackage{epsf}

\shortauthors{Bell 2002}
\shorttitle{Systematic Errors in UV-derived SFRs}

\newcommand{\ha}{{\rm H$\alpha$ }}

\newcommand{\hi}{{\rm H{\sc i} }}
\newcommand{\hans}{{\rm H$\alpha$}}
\newcommand{\hii}{{\rm H\,{\sc ii} }}
\newcommand{\hiins}{{\rm H\,{\sc ii}}}
\newcommand{\fir}{{\rm FIR }}

\newcommand{\fuv}{{\rm FUV }}
\newcommand{\fuvns}{{\rm FUV}}

\newcommand{\pegase}{{\sc P\'egase }}

\newcommand{\afuv}{{$A_{\rm FUV}$ }}
\newcommand{\afuvns}{{$A_{\rm FUV}$}}

\slugcomment{{\sc Accepted by the Astrophysical Journal: } Scheduled for 20 September 2002}

\begin{document}

%%%%% Added the \def\head{ lark.

\def\head{

\title{Dust-induced Systematic Errors in Ultraviolet-derived Star
Formation Rates} 

\author{Eric F. Bell}
\affil{Steward Observatory, University of Arizona, 933 N. Cherry Avenue, 
   Tucson, AZ 85721}
\email{ebell@as.arizona.edu}

\begin{abstract}
Rest-frame far-ultraviolet (FUV) luminosities form the 
`backbone' of our understanding of star formation at all cosmic epochs.
These luminosities are typically corrected for dust by
assuming that the tight relationship between the UV spectral slopes 
($\beta$) and the FUV attenuations (\afuvns) 
of starburst galaxies applies for all star-forming galaxies.
Data from seven independent UV experiments
demonstrates that quiescent, `normal' star-forming galaxies 
deviate substantially from the starburst galaxy $\beta$--\afuv
correlation, in the sense that normal galaxies are redder 
than starbursts.  Spatially resolved data for the 
Large Magellanic Cloud suggests that dust geometry and properties, 
coupled with a small contribution from older stellar populations, 
cause deviations from the starburst galaxy $\beta$--\afuv correlation.
Folding in data for starbursts and ultra-luminous infrared galaxies,
it is clear that neither rest-frame UV-optical colors nor 
UV/\ha significantly help in constraining the UV attenuation.
These results argue that the estimation 
of SF rates from rest-frame UV and optical data alone is 
subject to large (factors of at least a few) systematic uncertainties
because of dust, which {\it cannot} be reliably corrected for
using only UV/optical diagnostics.
\end{abstract}

\keywords{ultraviolet: galaxies --- dust, extinction --- 
galaxies: general --- galaxies: stellar content }
}%%%end head

\twocolumn[\head]

\section{Introduction} \label{sec:intro}

Understanding the star formation (SF) rates
of galaxies, at a variety of cosmic
epochs, is a topic of intense current interest
\citep[e.g.][]{yan99,blain99,haarsma00}.
Many SF rates are derived from highly dust-sensitive rest frame 
far-ultraviolet (FUV) luminosities \citep[e.g.][]{madau96,steidel99}.
In the local Universe, \citet{calzetti94,calzetti95} and \citet{meurer99}
found a tight correlation between ultraviolet (UV)
spectral slope $\beta$\footnote{$F_{\lambda} 
\propto \lambda^{\beta}$, where $F_{\lambda}$ is the flux per 
unit wavelength $\lambda$} and the
attenuation\footnote{Attenuation differs from extinction in 
that attenuation describes the amount of light lost because of dust at a given
wavelength in systems with complex star/dust geometries
where many classic methods for determining extinction, such as
color excesses, may not apply.} in the FUV (\afuvns), for a sample of  
in\-hom\-ogen\-eous\-ly-selected starburst galaxies.  
This correlation's low scatter 
requires a constant intrinsic value of $\beta \sim -2.5$
for young stellar populations \citep[e.g.][]{sb99}, coupled
with some regularities in the distribution and extinction properties
of dust \citep[e.g.][]{gordon97}.
Assuming that this
correlation holds for all galaxies at high redshift, this was used to 
correct the FUV flux for extinction in a statistical
sense \citep[see e.g.][and references therein]{adel00}.

However, recent work has called the universality of 
the $\beta$--\afuv correlation into question.  
Radiative transfer models predict a large scatter between
$\beta$ and \afuv \citep{clump2}.
Furthermore, both Large Magellanic Cloud (LMC)
\hii regions \citep{lmc} and ultra-luminous infrared galaxies
\citep[ULIRGs;][]{gold02} do not obey the starburst
correlation.  Tantalizingly, there are indications that
`normal', quiescent star-forming galaxies have less 
UV extinction than predicted by the Calzetti et al.\
relation \citep{buat02}.  Taken together, these issues
raise serious questions about the applicability of SF rates
derived from rest-frame UV light alone for non-starbursting 
galaxies.

In this paper, I investigate the relationship between
$\beta$ and \afuv for quiescent, `normal' star-forming galaxies
for the first time (to date, this correlation has been examined
directly for starbursts, ULIRGs and \hii regions only).
In \S \ref{sec:beta}, I 
demonstrate that normal galaxies deviate substantially 
from the starburst
galaxy $\beta$--\afuv correlation.  In \S \ref{sec:origin}, 
I use spatially-resolved
data for the LMC to investigate the origin of this effect.
In \S \ref{sec:alt}, I explore potential alternatives to $\beta$,
and briefly discuss the conclusions in \S \ref{sec:conc}. 
In the following, all UV and optical data are corrected
for galactic foreground reddening following \citet{sfd}, unless
stated otherwise.

\section{The $\beta$--\afuv correlation for normal galaxies} \label{sec:beta}

Sections \ref{subsec:deriv} and \ref{subsec:data} describe
the detailed derivation and uncertainties of $\beta$ and \afuv
estimates for normal galaxies: casual 
readers can skip ahead to \S \ref{subsec:results}.

\subsection{Calculation of $\beta$ and \afuv} \label{subsec:deriv}

Because most of the well-studied quiescent, 
`normal' star-forming galaxies greatly exceed
the $10\arcsec \times 20\arcsec$ aperture size of the 
{\it International Ultraviolet Explorer} (IUE), values of 
$\beta$ are calculated from multi-passband integrated and large-aperture
photometry, taking into account galactic foreground
extinction.  The UV spectral slope $\beta$ 
is estimated between 1500{\AA} and 2500{\AA}
for most galaxies (except for the LMC, 
where it is estimated between 1500{\AA} and 1900{\AA}).  Exploration
of both models \citep{sb99} and the IUE spectra of the central parts of 
seven `normal' galaxies
\citep{kinney93} indicates that there are no significant offsets
between values of $\beta$ determined between 1500{\AA} and 2500{\AA}
for normal galaxies
and those determined in other ways \citep[e.g. between 1200{\AA} and 
1900{\AA};][]{meurer99,adel00}.

The FUV attenuation, \afuvns, is estimated
by balancing the total, integrated FUV light (at wavelengths
between 1474{\AA} and 1650{\AA}, depending on experiment)
against the far-infrared (FIR) luminosity.  I use
the flux ratio method of \citet{fluxrat} to estimate
\afuvns, which accounts for different dust geometries
and stellar populations (assuming a young, $\sim$ 100 Myr old constant
SF rate population for starbursts, 
and an older, $\sim$ 10 Gyr old constant SF rate
population for normal galaxies).  A simple estimate which
assumes that young FUV-emitting stars completely
dominate the heating,
$A_{\rm Simple} = -2.5 \log_{10} L_{\rm FUV} / (L_{\rm FUV} + L_{\rm FIR})$,
where $L_{\rm FUV} = \nu f_{\nu}$ is evaluated near 1500{\AA} and 
$L_{\rm FIR}$ is an estimate of the total \fir luminosity,
overestimates flux ratio method-derived attenuations \afuv by $\la 10$\%
(because $A_{\rm Simple}$ does not account for the fact that some 
FIR is reprocessed light from older stars).  
If, for normal 
galaxies, reprocessed light from older stars dominates the FIR
luminosity, their \afuv should be interpreted as upper limits
(strengthening these conclusions).
The total \fir luminosity for all of the sample galaxies 
is estimated between $8-1000${\micron} by extrapolating
the observed $8-100${\micron} flux to longer wavelengths using 
a modified blackbody with an emissivity $\propto \lambda^{-1}$.  
Comparison of this
algorithm for estimating
$L_{\rm FIR}$ with $150-205${\micron} {\it Infrared Space Observatory} (ISO)
measurements suggests that the $L_{\rm FIR}$ are accurate for 
early-type spirals, but may underestimate the true $L_{\rm FIR}$
for starbursts and late-type spirals by $\sim$25\% 
\citep{calzetti00,popescu02}.  
Including the $L_{\rm FIR}$ from
cold dust will (somewhat counter-intuitively) move the 
starbursts by $\sim$0.2 mag to higher \afuvns,
but will affect the normal galaxies less (again, strengthening these 
conclusions).  These \afuv are directly comparable to the `IR excess' 
of e.g.\ \citet{meurer99} and the UV extinctions estimated at 
2000{\AA} by e.g.\ \citet{buat96} and \citet{buat02}.  

\subsection{The Data} \label{subsec:data}

\begin{deluxetable}{lc}
\tablewidth{144pt}
\tablecaption{Newly-derived UIT $A1 - B1$ colors \label{tab:data}}
\tablehead{\colhead{Galaxy Name} & \colhead{$A1 - B1$}  }
\startdata
M33 (NGC 598) & 0.70 $\pm$ 0.05 \\
M74 (NGC 628)& 0.33 $\pm$ 0.04 \\
M77 (NGC 1068)& 0.37 $\pm$ 0.07 \\
NGC 1317 & 0.39 $\pm$ 0.02 \\
NGC 2993 & 0.67 $\pm$ 0.05 \\
M81 (NGC 3031) & 0.22 $\pm$ 0.05 \\
UGC 6697 & 0.4 $\pm$ 0.1 \\
\enddata
\\ These colors have not been corrected \\ for 
foreground galactic extinction.
\end{deluxetable}

Accurate $A1 (\lambda_{\rm eff} \sim 2488${\AA})$ - 
B1 (\lambda_{\rm eff} \sim 1521${\AA}) colors
for 7 galaxies with {\it Ultraviolet
Imaging Telescope} \citep[UIT;][]{stecher97} data 
were derived using circular apertures with sizes designed to 
optimize $S/N$ while encompassing the majority of 
the galaxian light (Table \ref{tab:data}).  These colors were corrected to 
the effective wavelengths following the method of \citet{lmc}, and
$\beta$ is then derived using the color- and galactic
foreground extinction-corrected $A1-B1$ assuming 
$\beta = 0.4(A1-B1)/\log_{10}(\lambda_{{\rm eff},B1}/\lambda_{{\rm eff},A1})$.
To derive \afuvns, FUV $B1$ magnitudes and total FIR 
fluxes were taken from \citet{uit}, except for M33 (NGC 598): 
I have measured $B1 = 5.6 \pm 0.1$, 
and have adopted FIR fluxes from \citet{rice88}.  M81's $A1-B1$ colors
are for the star-forming outer regions only.  The colors and 
magnitudes are consistent to $\la$0.2 mag
with data from other UV satellites
\citep{uit}, and with different analyses of these data 
\citep[e.g.][]{marcum01}.

Galaxies with homogenized 10{\arcmin} aperture
{\it Orbiting Astronomical Observatory} (OAO) flux measurements  
at 1550{\AA}, 1910{\AA} and 2460{\AA}, or homogenized 
2.5{\arcmin} square aperture
{\it Astronomical Netherlands Satellite} (ANS) flux measurements  
at three or more of 1550{\AA}, 1800{\AA}, 2200{\AA} and 2500{\AA}, 
were taken from the compilation of \citet{rifatto2}.
Eight (twelve) galaxies with an observed OAO (ANS) magnitude at 1550{\AA} of 
$\leq 9.7$\,(12.2)
were selected, to ensure reasonable $S/N$.  Five galaxies
are common to both samples.
Values of $\beta$ are calculated using a simple
linear fit to the foreground extinction-corrected UV
magnitudes.  The OAO- and ANS-derived $\beta$ 
values are consistent to $\la 0.5$ for the
5 galaxies in common, suggesting errors $\la 0.35$ (as some 
of the discrepancies could be due to aperture mismatch).
Values of \afuv are calculated using
`total' UV magnitudes from \citet{rifatto3} and 
{\it Infrared Astronomical Satellite} (IRAS) fluxes from 
(in order of preference) \citet{rice88}, \citet{soifer89} or
\citet{moshir90}.  Comparison with UIT data showed that 
`total' UV magnitudes from \citet{rifatto3} derived
from large-aperture UV data are good to $\la$0.2 mag \citep{uit}.

The properties of the LMC are estimated using the images
presented by \citet{lmc}.  
Large aperture photometry of the LMC yields $\beta = -0.2 \pm 0.3$
and $A_{\rm FUV} = 0.54 \pm 0.21$, whereas totaling
the areas detected at $> 5 \sigma$ yields $\beta = -0.7 \pm 0.3$
and $A_{\rm FUV} = 0.48 \pm 0.13$.  These are quite consistent
with $\beta \sim -0.7$ from the D2B-Aura Satellite \citep{maucherat80}.
A 5{\arcmin} aperture around 30 Dor 
gives $\beta = -1.3 \pm 0.3$ and $A_{\rm FUV} = 1.4 \pm 0.2$ 
\citep[consistent with TD1 and ANS values for the area
of $\beta \sim -1.1$; ][]{maucherat80}.

As useful comparison samples, I include {\it i)} 
19 starbursts with UV spectra from 
the IUE in $10\arcsec \times 20\arcsec$ apertures, with
diameters smaller than 1.5{\arcmin} to minimize aperture effects
\citep{calzetti94,calzetti95,meurer99},
and {\it ii)} 7 ULIRGs with total UV fluxes and colors from {\it Hubble
Space Telescope} STIS data
\citep{gold02}.

\subsection{Results} \label{subsec:results}

\begin{figure}[tbh]
\vspace{-0.5cm}
\hspace{-0.5cm}
\epsfbox{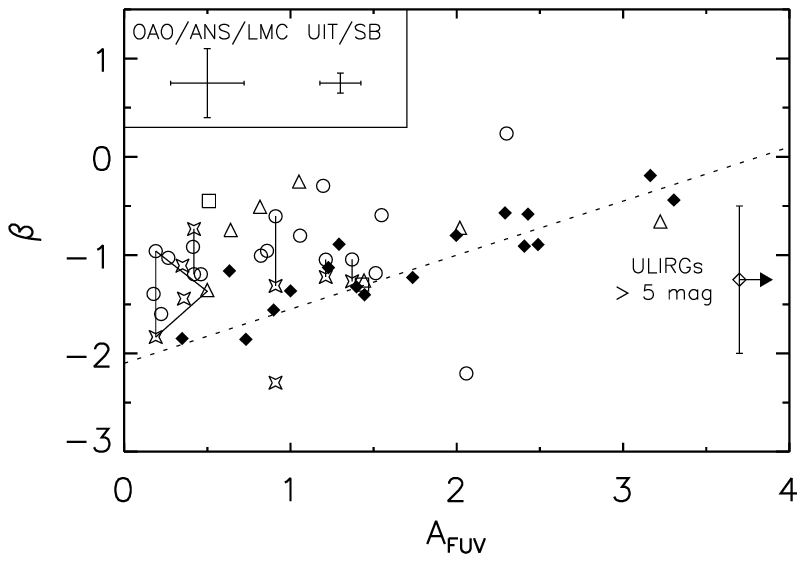}
\vspace{-0.5cm}
\caption{\label{fig:beta} 
UV spectral slope $\beta$ against attenuation at
$\sim 1500${\AA}, \afuvns.  UIT galaxies are plotted
as triangles, OAO galaxies as stars, ANS 
galaxies as circles, the LMC and 30 Dor are plotted
as squares, and starbursts are plotted as filled
diamonds.  ULIRGs (open diamond) typically have $A_{\rm FUV} \ga 5$ mag,
and `blue' $\beta$ values \protect\citep{gold02}.  
The dotted line shows a rough `by-eye' fit
to the starbursts.  Typical errors are depicted for 
OAO/ANS and the LMC, and for the UIT and starbursting galaxies.
Different measurements for the same galaxies are 
connected by solid lines.
}
\end{figure}

Fig.\ \ref{fig:beta} demonstrates that normal galaxies (open symbols)
have substantially redder UV spectral slopes, by $\Delta \beta \sim 1$,
than their starburst counterparts (filled diamonds) 
at a given \afuv (derived using FUV/FIR energy balance).
Furthermore, normal galaxies exhibit substantially
larger scatter than the starbursts.

This result is supported by detailed examination of the 
distribution of galaxies in the $\beta$--\afuv plane.
Focusing on the UIT galaxies (triangles), three starbursting galaxies
(M77, NGC 1317 and NGC 2993) lie on the Calzetti et al.\
relation, whereas the normal galaxies M33, M74, M81, and 
UGC 6697, do not.  The LMC lies redwards of the starburst
line, whereas 30 Dor, long held as an example of a `mini-starburst',
lies on the Calzetti et al.\ relation.
All of the OAO and ANS galaxies (stars and circles respectively) 
within $\Delta \beta \sim 0.3$
of the starburst relation are intensely forming
stars (NGC 4631, M94 and NGC 5248; NGC 4490 is
also peculiar).  In contrast, OAO and ANS galaxies
lying redwards of the starbursting relation are 
typically forming stars somewhat less intensely.
It is not, at this time, clear why NGC 4605 and
NGC 7331 lie bluewards of the Calzetti et al.\ relation.
IUE data also suggest a $\beta$ offset between normal
and starburst galaxies: \citet{kinney93}
show that spiral
galaxies are $\Delta \beta \sim 0.5 (1)$ redder than
starburst (blue compact and \hiins) galaxies, even
in the $10\arcsec \times 20\arcsec$ aperture of the IUE.

The ubiquity of the offset between starbursting
and normal galaxies as measured by 7 different experiments
(UIT, OAO, ANS, and IUE for normal galaxies; 
a sounding rocket, D2B-Aura and TD1 for the LMC)
argues strongly that the offset is real, and is not 
simply instrumental in origin.  This conclusion
agrees with \citet{buat02}, who found that applying the 
\citet{calzetti94} extinction law
to normal galaxies overestimated the UV extinction, using
an independent sample of galaxies without UV spectral slopes.

\section{Exploring the origins of the $\beta$--\afuv correlation} 
\label{sec:origin}

\begin{figure}[tbh]
\vspace{-0.5cm}
\hspace{-0.5cm}
\epsfxsize=\linewidth
\epsfbox{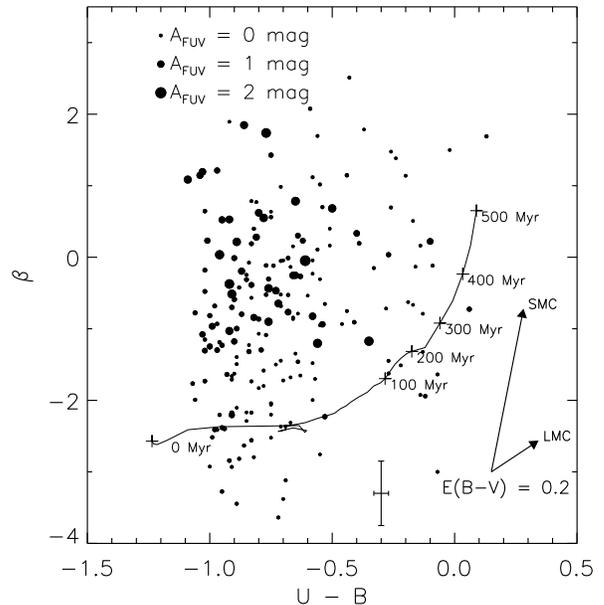}
\vspace{-0.4cm}
\caption{\label{fig:bica} 
UV spectral slope $\beta$ against $U - B$ color (a reasonable
age indicator) for a sample of 198 UV-bright stellar clusters and
associations in the LMC.  Symbols are coded by FUV attenuation:
larger symbols depict more highly attenuated clusters.  
Overplotted are the \pegase stellar population model colors
for a single burst with ages $<500$ Myr (Fioc \& Rocca-Volmerange, 
in preparation).  The effect of 
dust reddening, assuming a SMC bar-type or LMC-type dust screen,
is shown.  The dust vectors are displayed simply to give the 
reader some intuition about the effects of dust: radiative
transfer effects and/or extinction curve variations in the UV
make the prediction of the detailed effects of dust on a plot of this
type challenging \protect\citep{fluxrat,lmc}.
Typical errors are denoted by the error bar.
}
\end{figure}

It is reasonable to hypothesize that stellar population
and/or dust effects contribute to the
difference in behavior between normal and starburst galaxies
on the $\beta$--\afuv plane.
Stellar population models show that older {\it star-forming}
stellar populations are somewhat redder ($\Delta \beta \la 0.5$)
than younger star-forming populations
\citep[e.g.][]{sb99}.
Alternatively, radiative transfer models 
can easily generate relatively large changes in 
$\beta$ for only modest \afuv by appealing to different
star/dust geometries and/or extinction curves 
\citep{fluxrat,lmc}.

In order to test why normal galaxies have redder $\beta$ values
than starbursts, independent constraints on a galaxy's star
formation history are required (to allow splitting of age and dust
effects).  Independent age constraints are available
for stellar clusters and associations in the LMC (in the form of
$U - B$ optical color).  
Using $\sim 1\arcmin$ resolution multi-wavelength 
images \citep{lmc}, I have constructed 
5{\arcmin} (67 pc) aperture values of $\beta$ and \afuv for 
a sample of stellar clusters and associations 
from \citet{bica96}.  I choose 198 clusters and associations with $U < 12$ to
maximize the chances that the clusters dominate the aperture FUV flux.
Symbol size in Fig.\ \ref{fig:bica} reflects a cluster's 
UV attenuation, as estimated from FUV/FIR.  The colors
of a single burst stellar population and dust screen reddening
vectors are also shown.

It is clear that {\it only} young, unattenuated clusters have
`blue' $\beta \sim -2$ values.  Redder, $-1 \la \beta \la 1$
clusters tend to be {\it either} relatively young but
attenuated (the clusters with $U - B \sim -0.8$, but redder $\beta$ 
values) {\it or}  older and dust free (the clusters 
with $U - B \sim 0$).  

The balance between dust and old stellar
population effects can be constrained by considering the 
fraction of the total $U < 12$ LMC cluster UV luminosity 
which each population represents.  The young, unattenuated clusters
(defined to have $U - B \leq -0.5$ and $\beta \leq -1.2$, 1/3 of 
the clusters by number) represent
67\% of the summed $U < 12$ LMC cluster FUV 1500{\AA} 
luminosity.  The younger, attenuated
clusters (defined to have $U - B \leq -0.5$ and $\beta > -1.2$, half
of the clusters by number) 
represent 27\% of the FUV luminosity.  The older, unattenuated clusters
(defined to have $U - B > -0.5$, 1/6 of the clusters by number) 
have only 6\% of the FUV 
luminosity.  Noting that the clusters are selected to have 
$U < 12$ (and therefore are being selected to be reasonably
young) and possible differences between the cluster and field
SF histories, this result tentatively ascribes much of the 
observed `redness' of the LMC to dust effects:
older stellar populations tend to be UV-faint and do not 
affect the global $\beta$ estimate as significantly.  This 
interpretation is consistent with the detailed results of stellar
population modeling. \citet{sb99} show that the {\it maximum} 
possible offset $\Delta\beta$ between young 
and older {\it star-forming} stellar 
populations is $\sim$0.5, as, to first order, stars that are bright
enough to affect the FUV luminosity of a galaxy with even a small
amount of ongoing SF have very blue $\beta$ values.

One intriguing feature of Fig.\ \ref{fig:bica} is the population of 
seemingly unattenuated clusters
with $-1 \la \beta \la 1$ but blue $U - B$ colors.  While
the LMC sounding rocket images 
are photographic and relatively old \citep{smith87},
extensive checks against UIT, IUE, D2B-Aura, 
TD1 and ANS data have not indicated any significant 
problem with the calibration and $\beta$ values of the
LMC data \citep[see also][]{lmc}.  This argues that most
of these clusters do, in reality, have red $\beta$ but 
blue $U - B$ color with relatively little associated FIR.
Behavior of this kind can be generated by a large dust shell or 
extended screen geometry.  The UV light is attenuated and reddened
by the dust shell or screen, but the dust heating is spread
out over a much larger area of the sky than that subtended by 
the UV light from the cluster, and so the cluster appears to 
have high FUV/FIR.  

This can be explored more quantitatively by 
inspection of a simple model.  Assume that highly extincted
clusters ($\sim$ 1.5 mag, implying FIR/FUV$ = 3$) 
are attenuated by a
homogeneous shell of dust with a diameter of 5{\arcmin}, or 67 pc, on
the sky (this is a conservative 
assumption as the shells could be smaller).  Assume that a
0.3 mag attenuated cluster has a similar shell, 
only larger (so that the FIR/FUV$_{\rm total} = 3$, but FIR/FUV in the 
5{\arcmin} aperture is 1/3).  Assuming conservatively 
that the surface brightness is constant over the face of this 
dust shell, the area of the dust shell in the 0.3 
mag extinction case is 9 times larger than in the 1.5 mag
extinction case, so the radius is a factor of 3 larger, at
100 pc.  This shell radius is comparable
to the \hi thickness of the LMC, and is not an unreasonable separation 
between the young, FUV-bright stars and the dust which attenuates them
\cite{elmegreen01}.

While the picture is a little complex, Fig.\ \ref{fig:bica}
suggests that the LMC has a relatively red $\beta \sim -0.5$
primarily because of dust effects (from radiative
transfer and/or extinction curve variations), with 
a modest contribution from the light of older stellar populations.
This lends weight to an interpretation of 
the redder $\beta$ values of normal galaxies mostly in 
terms of dust, with a small contribution from SF histories.

\section{Exploring alternatives to $\beta$} \label{sec:alt}

\begin{figure}[tbh]
\vspace{-0.2cm}
\hspace{1.0cm}
\epsfxsize=7.5cm
\epsfbox{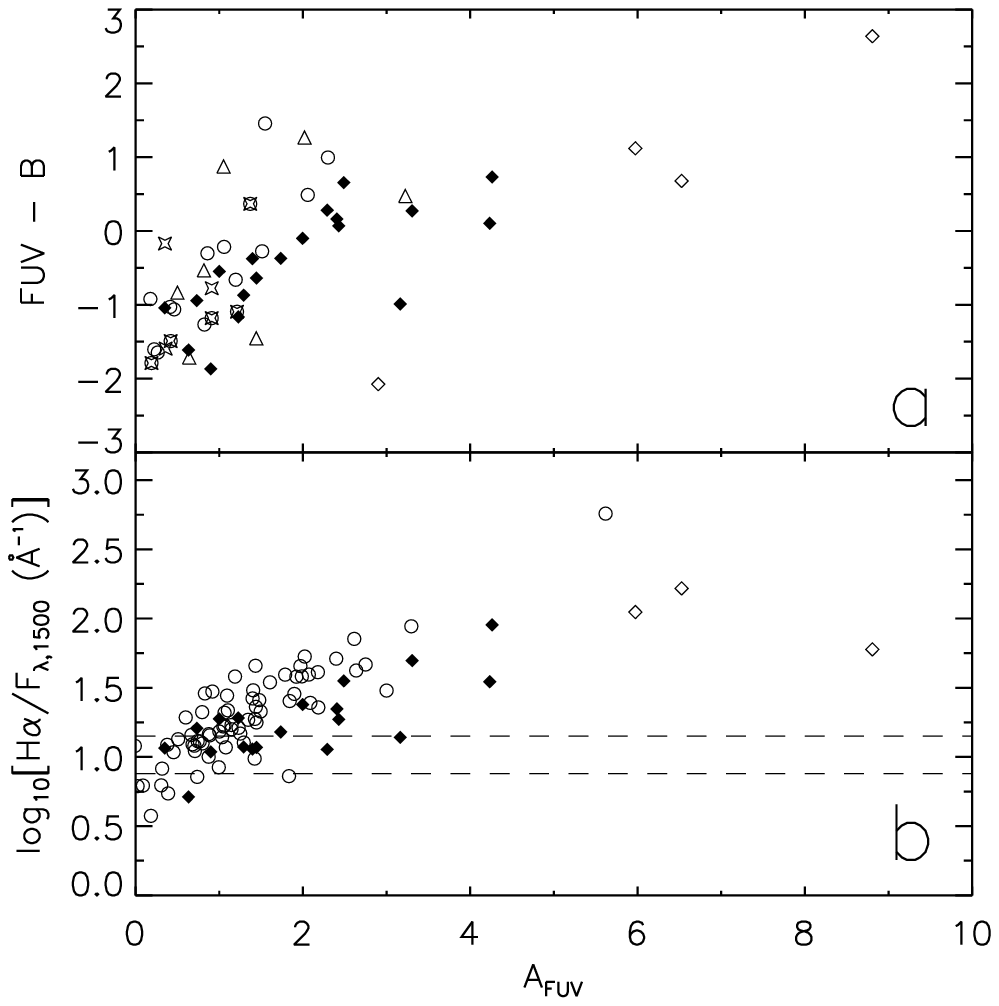}
\vspace{0.8cm}
\caption{\label{fig:alt} 
Alternatives to $\beta$.  Panel {\it a)} shows how FUV$ - B$ 
colors correlate with \afuvns. 
Symbols are as in Fig.\ \protect\ref{fig:beta}.
Panel {\it b)} shows \hans/\fuv against \afuvns.
Galaxies from \protect\citet{uit}
are shown as open circles.  \hans/\fuv are shown 
within the IUE or a 3.5\arcsec aperture for
starburst galaxies (filled diamonds) and ULIRGs (open diamonds)
respectively.  The two dashed lines depict the rough
dust-free value of \hans/\fuv \protect\citep{k98,sul00}.
}
\end{figure}

Fig.\ \ref{fig:beta} demonstrates that it is difficult
to estimate UV attenuations
on the basis of UV colors alone.  For example, a galaxy with 
$\beta \sim -1$ could have zero attenuation (if it is a 
relatively quiescent galaxy), or many magnitudes of 
attenuation (if it is a ULIRG).  
In Fig.\ \ref{fig:alt}, I examine two possible 
alternatives to $\beta$ which use only rest-frame
UV and optical data (and are therefore more easily
accessible to researchers wishing to determine SF rates
at high redshift).  

It is conceivable that FUV$ - B$, because of its longer
wavelength range, may be more robust to dust
radiative transfer effects than the UV
spectral slope $\beta$ (but would suffer more
acutely from the effects of older stellar populations). 
Indeed, Adelberger (private communication) reports that 
FUV$ - B$ tends to be a more robust dust indicator than $\beta$
for high redshift samples of galaxies. 
In panel {\it a)} of Fig.\ \ref{fig:alt}, I show
1550{\AA} FUV$ - B$ colors against \afuv 
for the UIT, OAO, ANS, ULIRG, and 
starburst galaxy samples.  Total FUV and $B$ magnitudes
were used for the UIT, OAO, ANS and ULIRG samples, whereas
the $B$ magnitudes of the starburst galaxies are roughly corrected
to the IUE aperture
following \citet{calzetti95}.  It is possible that 
FUV$ - B$ is a slightly more robust indicator than 
$\beta$, in terms of estimating \afuvns.  However, 
the scatter is enormous: at a given FUV$ - B$ color,
the range in \afuv is 3--5 magnitudes, or between 1 and 2
orders of magnitude.

\citet{buat02} suggest another potential extinction indicator:
\hans/\fuvns.  This indicator has the virtue that it is almost
independent of SF history \citep[although see][]{sul00}; however, 
it does depend on the relative distribution of dust around 
\hii regions compared to the dust around OB associations 
\citep[see e.g.][]{lmc}.
In panel {\it b)} of Fig.\ \ref{fig:alt}, I show integrated
\hans/\fuv for a sample of normal galaxies from \citet[filled circles]{uit}.
\hans/\fuv for starburst galaxies within the IUE aperture
\citep{calzetti94}, and 3.5{\arcsec} aperture values for 
ULIRGs \citep{gold02,wu98} are also shown.
In agreement with \citet{buat02}, I find that 
there is a scattered correlation 
between \hans/\fuv and \afuvns; however, the scatter is a 
challenge to its usefulness.  For example, 
at \hans/\fuv$ \sim 10${\AA}$^{-1}$, $0\,{\rm mag} 
\la A_{\rm FUV} \la 3\,{\rm mag}$, and at \hans/\fuv$ \sim 100${\AA}$^{-1}$, 
$A_{\rm FUV}  \ga 3\,{\rm mag}$.  

Importantly, at a given \afuvns, 
starbursts and ULIRGs tend to have bluer $\beta$ values, bluer FUV$ - B$
colors, and lower \hans/\fuv than normal galaxies.
Thus, SF rates derived from rest frame UV data, even
analyzed in conjunction with rest frame optical data, suffer from 
{\it systematic} uncertainties of at least factors of a few.

Because of the sample's inhomogeneous selection, it is 
difficult to reproduce the UV-bright selection imposed
on distant samples of star-forming galaxies, and thus
to make detailed predictions about the degree of systematic
error introduced in the determination of SF rates at high redshift.  
However, since the physical nature of distant star-forming 
galaxies is still poorly understood, these results
argue that UV-derived SF rates of high redshift galaxies
should, at best, be viewed with caution.

\section{Conclusions} \label{sec:conc}

I have used data from 7 independent UV experiments
to show that quiescent, `normal' star-forming galaxies 
have substantially redder UV spectral slopes $\beta$ at
a given \afuv than starbursting galaxies.
Using spatially resolved data for the LMC, 
I suggest that dust geometry and properties, coupled with 
a small contribution from older stellar populations, 
contribute to deviations from the
starburst galaxy $\beta$--\afuv correlation.
Neither rest frame UV-optical colors nor 
UV/\ha significantly help in constraining the UV attenuation.
These results argue that the estimation 
of SF rates from rest-frame UV and optical data alone is 
subject to large (factors of at least a few) systematic uncertainties
because of dust, which {\it cannot} be reliably corrected for
using only UV/optical diagnostics.

\acknowledgements

I wish to thank Kurt Adelberger, Daniela Calzetti, Betsy Barton Gillespie,
Karl Gordon, Rob Kennicutt, Mark Seibert, Richard Tuffs, and Dennis Zaritsky
for useful discussions and suggestions.  Special thanks go to 
the anonymous referee for their careful suggestions.
This work was supported by NASA grant NAG5-8426 and NSF grant AST-9900789.  
This work made use of NASA's Astrophysics Data System 
Bibliographic Services, and the NASA/IPAC Extragalactic Database 
(NED) which is operated by the Jet Propulsion Laboratory, 
California Institute of Technology, under contract with the 
National Aeronautics and Space Administration.

\end{document}